# Machine Learning Forecasting of Active Nematics

Zhengyang Zhou[1], Chaitanya Joshi[2], Ruoshi Liu[2], Michael M. Norton[2], Linnea Lemma[2], Zvonimir Dogic[3], Michael F. Hagan[2], Seth Fraden[2], Pengyu Hong[1]

**Abstract**: Active nematics are a class of far-from-equilibrium materials characterized by local orientational order of force-generating, anisotropic constitutes. Traditional methods for predicting the dynamics of active nematics rely on hydrodynamic models, which accurately describe idealized flows and many of the steady-state properties, but do not capture certain detailed dynamics of experimental active nematics. We have developed a deep learning approach that uses a Convolutional Long-Short-Term-Memory (ConvLSTM) algorithm to automatically learn and forecast the dynamics of active nematics. We demonstrate our purely data-driven approach on experiments of 2D unconfined active nematics of extensile microtubule bundles, as well as on data from numerical simulations of active nematics.

## Introduction

Active nematics are far-from-equilibrium materials with local orientational order, whose anisotropic constituents consume energy at the particle scale to generate forces and motions [1-7]. Being driven away from equilibrium, active nematics have the potential to transform materials science by enabling a new class of materials with capabilities currently found only within living organisms. Uniformly aligned active nematics are inherently unstable. Instead, in steady state they exhibit chaotic turbulent-like dynamics that lack long-range order or the ability to drive net material transport (e.g.[1, 5, 8-19]). Long-term practical applications of active nematics require developing adaptive control strategies that transform their turbulent dynamics into stable flows that generate productive work [20]. The recent development of light-controlled molecular motors paves the way toward this possibility [21, 22]. A first step toward this long-term goal is the ability to forecast the temporal evolution of the active nematic dynamics.

One plausible path toward forecasting active nematic dynamics involves developing a quantitative predictive theoretical model of active nematics. In such a case, one could use the experimentally measured director and velocity fields as initial conditions and forecast the subsequent dynamics. However, this path is fraught with significant challenges. Existing continuum models capture many structural and dynamical features of active nematics in a statistical sense [1, 2, 4, 5, 7, 12, 14, 19, 23-38], but they fail to account for certain physical aspects of the experimental dynamics [39]. For example, in existing continuum theories defects can propagate through molecular reorientation of the director field, while in cytoskeletal active nematics, defects avoid crossing material lines defined by the director field because their motion is constrained by the long spatial extent of the constituent fibers [39]. Furthermore, theoretical models of active nematics require numerous input parameters, such as the magnitude of the active stress and the nematic elasticity. These are not known a priori and are challenging to measure experimentally.

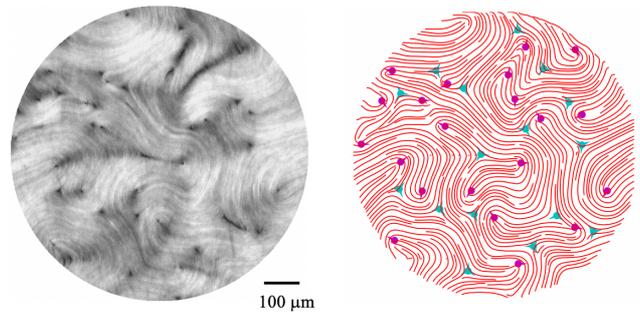

**Figure 1**. Example of an unconfined 2D active nematic system consisting of extensile microtubules. **Left**: Image of the optical retardance obtained using a PolScope. A circular view of interest is set in this study. **Right**: The corresponding orientation field (red lines), +½ defects (red disk with line indicating orientation), and –½ defect (blue tricuspoids). See the Data subsection in the main text for explanations of orientation field visualization and defect detection.

To overcome these challenges, we describe an alternate path toward forecasting active nematic dynamics that involves Deep Learning (DL) [40]. This method does not require any knowledge of the underlying physics, and only leverages the symmetries of the system. Our DL framework automatically constructs a deep neural network that capture the dynamics of an experimental active nematic system. Our model forecasts future movements for time scales long enough to capture key physical events such as defect nucleation and annihilation. This is an essential first step toward rational design of devices that harness the tantalizing capabilities of active nematics, and provides the basis for a closed-loop feedback control strategy. More broadly, by demonstrating the possibility of forecasting the dynamics of a complex active material, our results and those of a complementary independent work [41] extend the applications of machine learning in a significant new direction.

Deep Learning, a subfield of Machine Learning, has transformed a broad range of scientific disciplines by enabling understanding, exploring, and interpreting large-scale datasets. Typical DL models are deep artificial neural networks comprised of many layers of simple non-linear neurons. Each layer learns to transform its input (starting with the raw input) into a more abstract and composite representation. This allows very complex functions to be effectively decomposed and learned if the models have enough layers. The key aspect of DL is that these representations are not designed by human experts; instead, they are automatically learned from raw data using a general-purpose

1. Computer Science, Brandeis University
2. Physics, Brandeis University
3. Physics, University of California, Santa Barbara
\* Project github: https://github.com/zhengyjo/Machine-Learning-Forecasting-of-Active-Nematics-/tree/master/unconfined_orientation

learning procedure. The DL process can disentangle these representations and decide their places in a model to optimize the model's performance. Long-Short-Term-Memory (LSTM) networks are one DL model family [42], suitable for classifying, processing, and making predictions based on time series data. A ConvLSTM enables temporal-spatial modeling [43], by incorporating convolution operations into the LSTM to extract spatial information from inputs [44]. Our forecasting model is based on a ConvLSTM model, and uses eight consecutive frames of an active nematics to predict its future movements. Crucially, we demonstrate that the model can forecast key events in the dynamics, and could thus provide predictions for a closed-loop feedback control strategy.

## Methods

### Data

*Experimental data:* The training dataset contained five videos, each consisting of retardance and orientation fields, produced by PolScope microscopy of 2D active nematics of extensile microtubule bundles [11, 38, 45]. The time interval between two consecutive frames was two seconds. We observed in the experiments that crucial dynamical events like defect nucleation occurred within 8 video frames. Hence, we divided each video into sub-sequences of 16 consecutive frames (the first 8 frames as the input to the model and the last 8 frames as the outputs of the model). Two consecutive sub-sequences have 6 frames in overlap (i.e., the last 6 frames of a sub-sequence are the first 6 frames of the sub-sequence succeeding it). Based on the limited amount of available data, we separated the sub-sequences into two groups with a size ratio of ~10:1 for training and testing subsets, respectively. In particular, we used all the available videos to create 4,000 sub-sequences to train the model and another 400 to test it. The training sub-sequences came from the early segments of the videos and the test sub-sequences came from the late segments of the videos, with no frame overlap between the training set and the test set. A total of ~44,000 video frames were used in this study. With larger data sets, the percentage of sub-sequences needed for training would decrease; for example, with ten times more sub-sequences one could use 50% for training and 50% for testing. In this way, the percentage used for testing would increase significantly and the accuracy could be expected to also rise because of the increased number of sub-sequences used for training.

The original size of each frame was 1040×1040 pixels. We divided each frame into four 520×520 views, and then downsized each view into 128×128 pixels. The size of a typical topological defect was about 7×7 pixels in the downsized views. A circular view of interest was used in this study. The orientation ground truth was measured by a PolScope [45]. The visualizations of the nematic were made by integrating along the orientation vectors, while accounting for the nematic symmetry. The orientation field of a typical 2D active nematic in a chaotic steady state is continuous almost everywhere, except at few isolated points known as topological defects – the comet shaped +½ defects and the trefoil-like –½ defects (highlighted in the visualizations). The defect ground truth was obtained by our defect detection method (see details in the APPENDIX).

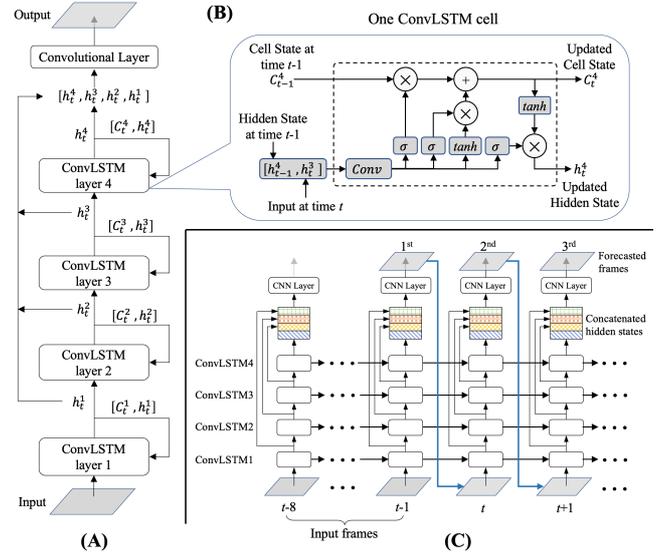

**Figure 2**. The stacked ConvLSTM model for forecasting active nematic movements. (**A**) The model contains four ConvLSTMs that are stacked together to capture relatively long-range temporal dependencies. $C_t^i$ encodes the long term memory and represents the state of the LSTM cell in *i*-th layer at the *t*-th frame. $h_t^i$ encodes the short term memory (i.e., the hidden state also known as output) of the LSTM cell in the *i*-th layer at the *t*-th frame. The outputs ($h_t^1, h_t^2, h_t^3$, and $h_t^4$) of all four ConvLSTMs are channel-wise concatenated before feeding to a 1×1 convolutional (CNN) layer to produce a forecasted frame. (**B**) Detailed schematic of one ConvLSTM cell. The input to the cell at time *t* is concatenated with its hidden state at time *t*-1, and then is passed through a convolutional (*Conv*) layer. The symbol $\sigma$ indicates the Sigmoid function. Both $\otimes$ and $\oplus$ are element-wise operators. (**C**) In action, the model is unrolled along the time axis to map an input sequence (from time *t*-8 to *t*-1) to an output one (1st, 2nd, 3rd, …). The forecasted frame at one time point is treated as the input of the next time point.

*Simulation data*: Simulation data was obtained by numerically solving the hydrodynamic equations for a simplified active nematics model (see the APPENDIX). A total of 9,900 frames of simulation data were used, with 9,000 frames used for training and the remaining 900 frames used for testing. The frames had a box size of 200×200 and were separated in time by $\Delta t = 10$, where both lengths and times are in non-dimensional simulation units. The solutions were computed on a 128×128 grid. In total, we compiled 900 training sequences and 90 test sequences.

Due to the nematic symmetry of the system, the range of an orientation vector is between [0, 180°]. A vector along 0° is the same as a direction vector of 180°, although their orientation degrees differ. Hence, we map the orientation angle $q$ into a nematic order parameter field as $[Q_{xx}, Q_{xy}] = [\cos^2(q) - ½, \cos(q) \cdot \sin(q)]$. To increase model robustness, we performed data augmentation by creating additional data sets via random rotations of the circular view of interest.

### Stacked ConvLSTM for forecasting active nematic movements

Figure 2 shows the architecture of our DL model for forecasting active nematics dynamics. It stacks four ConvLSTM layers, which allows the model to capture long-range temporal dependencies. The sizes of the hidden states in the stacked ConvLSTM layers are 32, 64, 64, and 128, respectively. Their corresponding convolutional kernel sizes are 7×7, 5×5, 5×5, and 5×5, respectively. The hidden states of all ConvLSTM cells are concatenated before being passed to a convolution layer to produce one output frame, which effectively forms a densely connected layer [46]. When used to forecast movements, an output frame (i.e., one prediction) is fed back as the next input to the bottom ConvLSTM in the model to produce the next predicted frame. We trained the model to take eight frames in history as the input and forecast eight future frames, by minimizing the mean-squared-error (MSE) loss. The model could have been trained with more than eight input frames, but this would require substantially more memory. In practice, the trained model can then be used to forecast more than eight future frames (Fig. 4, 5, and 9), though the accuracy of the model decays with number of frames into the future. The model was implemented using Pytorch [47]. The Adam training algorithm [48] was used with a learning rate of 0.001 and a step decay rate of 0.5 every 30 epochs.

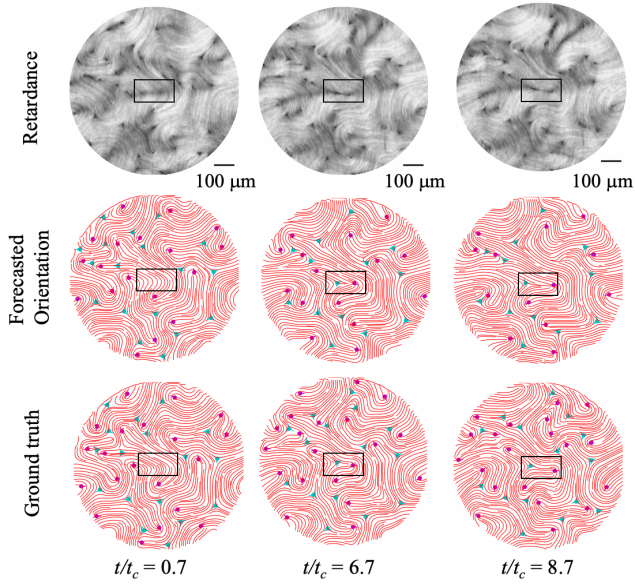

**Figure 3.** A example of active nematic dynamics forecasting, with a nucleation event highlighted in the boxes. The columns from left to right correspond to $t/t_c = $ 0.7, 6.7 and 8.7, respectively, where $t_c$ is the characteristic time for a +½ defect to move a distance equal to its core size. ***Top row***: Optical retardance images from the experiment. ***Middle row***: The forecasted orientation fields. ***Bottom row***: The ground truth of the orientation fields. The blue tricuspoids and red dishes (lines for orientations) indicate the −½ and +½ defects, respectively. The rectangles in the middle row highlight a pair of predicted defect nucleation events, which correspond to the ground-truth observation of nucleation events highlighted in the bottom row.

## Experiment Results

### An example of forecasting

Figure 3 shows an example of active nematic movement forecasting. Nematics typically contain localized regions of orientational disorder known as defects. There are two topological defect types (+½ and −½), which are defined by their topological winding number [49, 50]. The +½ defect is motile, and defects play a crucial role in the overall dynamics of active nematics. We therefore mark the defects in the ground truth and the forecasting results.

### Accuracy of forecasting

We characterize the forecasting fidelity ($F$) by comparing the predicted nematic director field and the ground truth, $F = \langle \cos(2\gamma) \rangle$, where $\gamma \in [0, \pi/2]$ is the absolute angular difference between a location in a ground truth frame and its prediction, and the $\langle \; \rangle$ operation denotes an average over test data at a given timepoint. $t_c$ is the characteristic time it takes for a +½ defect to move a distance equal to its core size (Appendix Fig. A1). We elaborate further on defects in the next section. In the simulations, this time scale is proportional to the ratio of the viscosity and activity (i.e., $t_c \propto \eta/\alpha$) [24]. From here on all the times are in units of the characteristic time, $t_c$.

The model performs well in forecasting early time frames but decays over time (Fig. 4). The error increases faster in the regions further away from the center. This spatial dependence of uncertainties in forecasting arises mainly because materials outside the circular view of interest, whose motions cannot be directly forecasted by the model, continually move into the view of interest.

### Predicting defect events

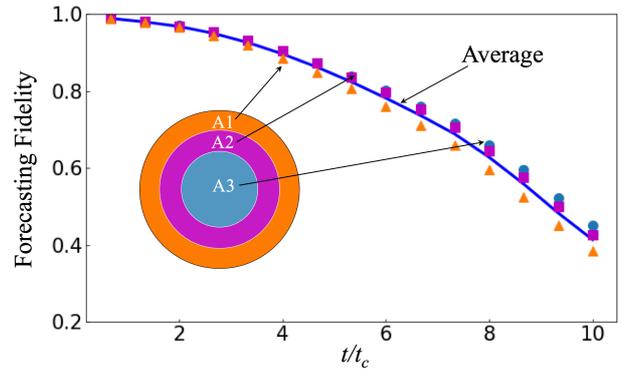

**Figure 4.** The forecasting fidelity of experimental data as a function of time in units of the characteristic time, $t_c$. The thick blue line shows the overall forecasting fidelity of the whole view, while the symbols show the spatial distribution of the forecasting fidelity. The colors of the symbols correspond to the colors of the regions (A1 – orange ring, A2 – purple ring, and A3 – the blue region) in the circular doman of the data, as indicated in the inset. The rings A1 and A2 have a width of 104 μm while the disc A3 has a radius of 208 μm.

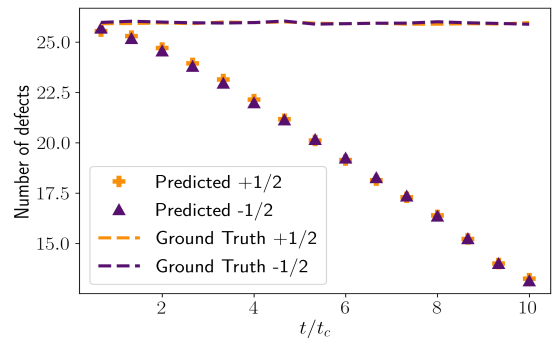

**Figure 5.** The overall performance of defect forecasting on the experimental data. Plotted are the number of defects in the viewed region that are detected in the data and forecasted by the model as a function of time, in units of the characteristic time, $t_c$. An analogous plot is shown for simulation data in Appendix Fig. A2.

Defect generation and motions are the critical physical processes that underlie the turbulent-like dynamics of active nematics. Therefore, in evaluating the performance of the model, we have focused on its ability to describe defect behaviors. In active nematics, the +½ defects are motile due to the active forces, whereas the –½ defects, due to their three-fold symmetry, do not experience a net force and are stationary on average. In our samples, the average speed of microtubules in the sample is about 1 μm/sec. The +½ defects move with a speed of ~4 μm/sec and have a mean diameter of ~12 μm (see APPENDIX); hence $t_c \approx 12/4 = 3$ sec. The defects stay in frame for ~50 seconds (~$16t_c$) before either annihilating (explained below) or moving outside of the frame. The framerate is 2 sec/frame, so 1 frame ≈ $0.7t_c$. Our model forecasts the number of defects in the system with an initial accuracy of ~98% for both +½ and –½ defects in the $0.7t_c$ forecasted frame. The accuracies drop gradually to ~50% for both +½ and –½ defects in the $10t_c$ (Figure 5).

Notably, this model can forecast key defect events, such as nucleation (Figure 6, movie S1), annihilation (Figure 7, movie S2), and splitting (Figure 8, movie S3). Nucleation occurs when a region with uniform nematic order undergoes bending due to active forces to form one +½ defect and one –½ defect (Figure 6, movie S1). Annihilation occurs when one +½ defect and one –½ defect combine to create a uniform nematic locally (Figure 7, movie S2). Splitting occurs when the nematic near a +½ defect itself undergoes bending to result in two +½

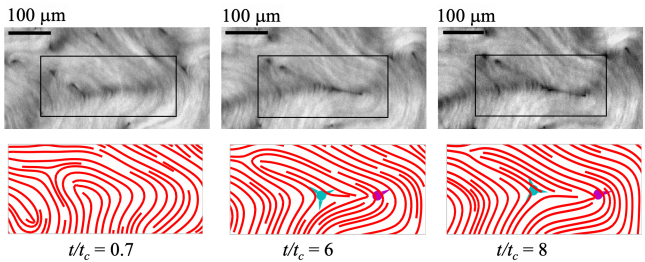

$t/t_c = 0.7$    $t/t_c = 6$    $t/t_c = 8$

**Figure 6**. Forecasting defect nucleation. The columns (from left to right) are regions in the forecasted frames at $t/t_c$ = 0.7, 6 and 8, respectively. The first row shows the ground truth retardance images. The second row shows the forecasted orientation fields corresponding to the rectangular regions in the first row. The blue tricuspoids and red dishes (lines for orientations) indicate the –½ and +½ defects of interest, respectively. See movie S1 in the APPENDIX for details.

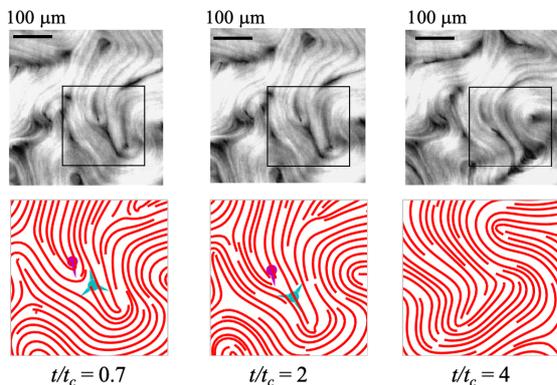

$t/t_c = 0.7$    $t/t_c = 2$    $t/t_c = 4$

**Figure 7**. Forecasting defect annihilation. The columns (from left to right) are regions in the data and forecasted frames at $t/t_c$ = 0.7, 2, and 4, respectively. The top row shows the ground truth retardance images. The bottom row shows the forecasted orientation fields corresponding to the rectangle regions in the first row. The blue tricuspoids and red dishes (lines for orientations) indicate the –½ and +½ defects of interest, respectively. See movie S2 in the APPENDIX for details.

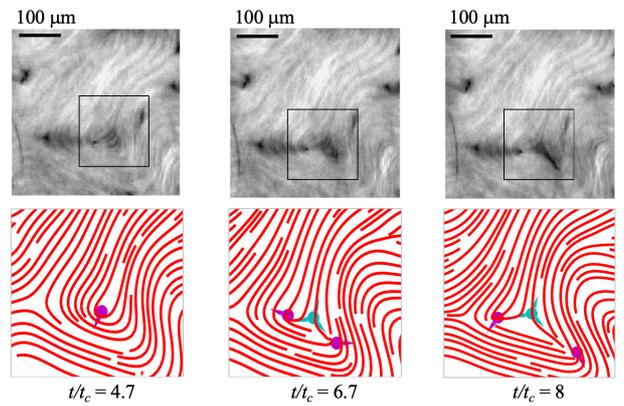

$t/t_c = 4.7$    $t/t_c = 6.7$    $t/t_c = 8$

**Figure 8**. Forecasting defect splitting. A +½ defect splits into two. The columns (from left to right) are regions in the data and forecasted frames at $t/t_c$= 4.7, 6.7, and 8, respectively. The top row shows the regions in the ground truth retardance images. The bottom row shows the forecasted orientation fields corresponding to the rectangle regions in the first row. The blue tricuspoids and red dishes (lines for orientations) indicate the –½ and +½ defects of interest, respectively. See movie S3 in the APPENDIX for details.

defects and one –½ defect (Figure 8, movie S3). The corresponding movies of these defect events are included in the APPENDIX.

**Simulation Results**

Finally, to test the generality of the DL framework for predicting active nematics dynamics, we tested the same approach on numerically simulated data of an active nematic. We used a minimal nemato-hydrodynamical model (see Methods and the APPENDIX) with parameters set to simulate active nematics dynamics in the turbulent regime. We numerically solved the equations using a multigrid solver, and calculated the orientation field ($Q_{xx}$ and $Q_{xy}$) from the simulation trajectories in the steady state. We trained our DL model using the same approach as described for the experimental data (Methods).

Similar to the experimental results, the forecasting fidelity of the DL model for the simulation data decreases as a function of time (Figure 9). To enable semi-quantitative comparison between forecasting experimental and simulation dynamics, we rescale the temporal evolution with the characteristic time of the corresponding dataset. For the simulation set, we find 1 frame ≈ $1.4t_c$. The +½ defects move at about 5 units per frame and survive about 25 frames (~$35t_c$) before annihilating. To the extent that the space and timescales match between experiments in theory, we see that the forecasting fidelity for the simulation data is somewhat higher, and remains accurate over a longer timescale, in comparison to the experimental data. Similarly, the prediction of defect density is accurate out to longer timescales for the simulation data (see Appendix Fig. A2). This performance increase can be understood at least in part because the simulation data is accurate to within roundoff error and is noise-free.

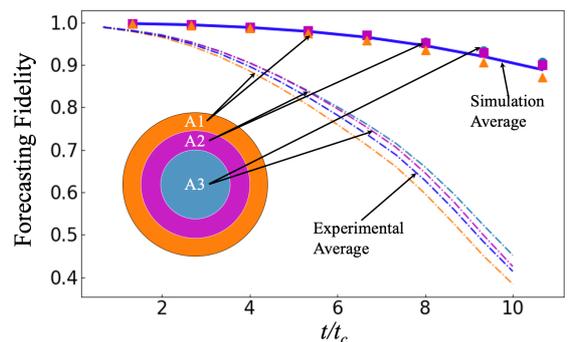

**Figure 9**. Forecasting fidelity of computational active nematic dynamics (solid lines) and on experimental data from Fig. 4 for comparison (dashed lines) as a function of time normalized by the characteristic time, $t/t_c$. The thick blue line shows the overall forecasting fidelity of the whole region, while the symbols show the spatial distribution of the forecasting fidelity. The colors of the symbols correspond to the colors of the regions (A1 – orange ring, A2 – purple ring, and A3 –inner blue region) in the circular doman of the data, as indicated in the inset.

**Discussion and Conclusions**

We present a data-driven approach to construct and train a ConvLSTM-based deep learning model to forecast the movements of 2D active nematics. The trained model is capable of forecasting key defect events such as creation, annihilation, and splitting, as validated against experiment. To test the generality of the approach, we have also shown its applicability to data from a hydrodynamic simulation. In a future study, we plan to characterize its performance on simulated data by varying the hydrodynamic parameters or adding noise.

Our model only relies on the symmetry of the system, and hence the structure of our model can also be applied to other kinds of anisotropic soft matter, such as colloidal liquid crystals, polar active matter self-propelled rods (e.g. [6, 51, 52]), and active-nematics-based vesicles and emulsions (e.g. [11, 53-55]). When applied to a new system, the model may need to be retrained using the data collected from the new system, and the size of the model may also need to be adjusted. It is possible that transfer learning can be used to quickly adapt a model trained in one system to a new system.

While the forecasting accuracy decays over time, this work demonstrates that deep learning is a viable approach for forecasting details of the turbulent dynamics of active nematics over time intervals corresponding to the displacement of a defect by several defect diameters. This marks an essential step toward the ability to implement feedback control strategies to generate desirable behaviors in active nematic systems. We found that the forecasting errors first accumulate in the following two types of regions: (1) local areas around defects (Figure 10 and movie S4); (2) boundary regions where uncertainty arises due to materials outside of the viewing area crossing the boundary into the viewing area (Figures 4 and 10, movie S4). These errors subsequently propagate to other areas in the later

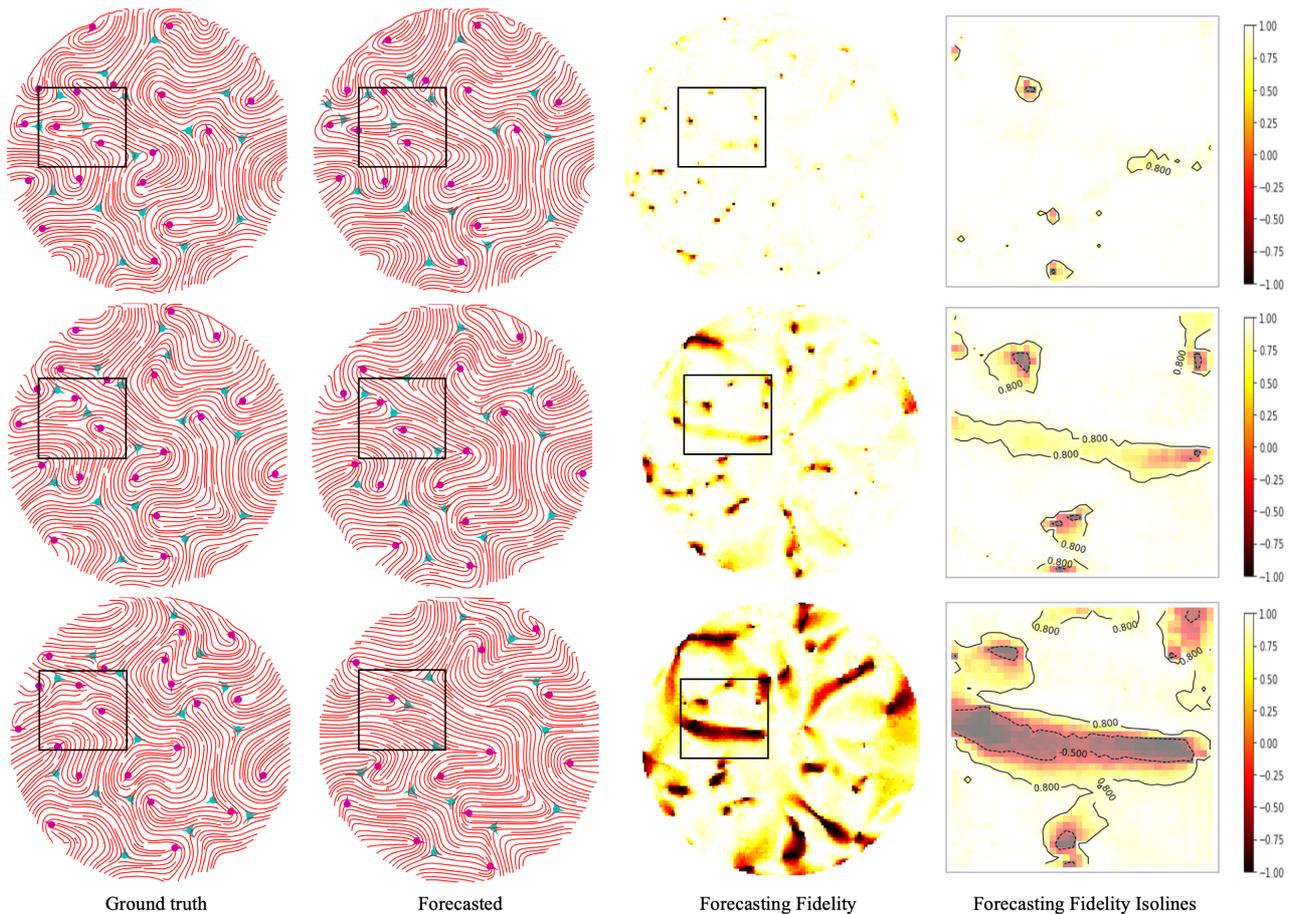

Ground truth     Forecasted     Forecasting Fidelity     Forecasting Fidelity Isolines

**Figure 10**. An example of the spatiotemporal accumulation and propogation of errors in forecasting experimental active nematics. The rows (from top to bottom) are at $t/t_c$ = 3.3, 6.7 and 10, respectively. The first column shows the ground truth orientation fields. The second column shows the forecasted orientation fields. The blue tricuspoids and the red dishes (with lines for orientations) indicate the −½ and +½ defects, respectively. The third column shows the corresponding spatial distributions of the forecasting fiedlity. The last column shows the contour plots of the corresponding box areas in column 3, in which the dark red areas surrounded by the solid line has forecasting fidelity less than 0.8 and the areas surrounded by the dashed lines have forecasting fidelity less than -0.5. See movie S4 in the APPENDIX for details.

forecasted frames. Identification of these error sources paves the way for improving data-driven forecasting algorithms.

Recent years have seen the emergence of machine learning approaches to study active matter systems (Ref. [56]). Thus far, efforts have focused on identification and tracking of particles (e.g. [57, 58]), classifying the collective dynamics of flocks/swarms (e.g. [59-63]), or the automatic identification of different dynamical phases (e.g. [64]). Most closely related to our objective, machine learning has been applied to train a relatively simple model to predict the simulated dynamics of chaotic systems produced by a computational model [65]. In comparison, our work demonstrates that machine learning can be used to train complicated models to forecast complex dynamics in experiments. While we were in the final stages of completing this manuscript, a related preprint appeared, which also describes a deep learning approach to forecast the dynamics of active nematics [41]. In comparison to their approach, our model stacks multiple ConvLTSMs with a larger convolutional view followed by dense connections, which potentially could capture large-scale temporal-spatial dependencies in active nematics motions. We empirically found this architecture to be useful for handling the chaotic nature of active nematics dynamics in our experiments. In addition, our model is purely data-driven and does not rely on knowledge of the topology of the defects or the elastic energy structure of the material. This feature avoids potential errors and uncertainties associated with using the algorithmically predicted defects in refining the predicted frames. However, we note that the total number of defects predicted by our algorithm for the experiments decays over time (see Fig. 5), whereas it does not in the predictions from the algorithm of Ref. [41]. One possible reason for this difference in performance is that Ref. [41] includes a model of the nematic in their sharpening algorithm, whereas our model is purely data-driven. This difference between the two approaches highlights that the model architectures and training protocols may be tuned to optimize particular performance goals. We hypothesize that the decay in the predicted number of defects is mainly due to two factors. First, we do not have enough training data to cover some local dynamics scenarios, which cause the forecasting errors to emerge and accumulate in local areas and then propagate into neighboring regions over time (Figure 10). Second, the nematic director outside of our model's circular view is artificially treated as horizontal, which produces biases that spread into the view of the model.

In the future, we plan to increase the forecasting accuracy of our approach by (a) improving the architecture of our model to address the hypothesized error sources; and (b) collecting more training data under a variety of experimental conditions to further improve the robustness and generalizability of the trained model. Our goal is to develop a model that can provide reliable guidelines for accurately controlling the behaviors of an active nematic system in an experiment.

**Conflicts of interest**
There are no conflicts to declare.

**Acknowledgements**
This work was supported by NSF DMR-MRSEC 2011486 and NSF DMR-1420382, NSF-DMR-1810077, NSF DMR-1855914, and NSF OAC 1920147. We also acknowledge computational support from NSF XSEDE computing resources allocation TG-MCB090163 (Stampede) and the Brandeis HPCC which is partially supported by DMR-MRSEC 2011486.

**Notes and references**

## APPENDIX

**Experiment procedure:** The active nematics were prepared following previously published protocols [11, 15]. Briefly, microtubules were polymerized from tubulin purified from bovine brain [66] at a concentration of 8 mg/mL in the presence of GMPCPP to stabilize the MT lengths. A truncated kinesin-1 motor proteins (K401-BCCP-HIS) was purified from E. Coli using immobilized metal affinity chromatography [67]. Motor clusters were formed by mixing 5 μL biotinylated kinesin (0.7 mg/mL) with 5.7 μL of streptavidin (0.34 mg/mL) and incubating for 30 min. The active mixture was composed of polyethylene glycol (0.8% w/v 20 kDa) as a depletion agent, phosphoenol pyruvate (26 mM), pyruvate kinase/lactic dehydrogenase (PK/LDH) and ATP (1.4 mM) as an energy regeneration system, and glucose (6.7 mg/mL), glucose oxidase (0.08 mg/mL), glucose catalase (0.4 mg/mL) and Trolox (2 mM) as an oxygen scavenging system. All components were mixed in M2B butter (80 mM PIPES, pH 6.8, 1mM EGTA, 2 mM MgCl2). Kinesin motor clusters (0.017 mg/mL K401) and microtubules (1.6 mg/mL) were added to the active mixture just before the experiment. A flow chamber 18 x 3 x 0.06 mm was made using double sided tape sandwiched between two glass slides. One slide was treated to be hydrophobic using Aquapel. The other slide was passivated with an acrylamide coating [68]. To form an active nematic, oil (HFE 7500) stabilized with a fluorosurfactant (1.8% w/v, RAN Biotech) was flowed in to fill the whole chamber. Then the aqueous active mixture was flowed in while wicking out the oil, leaving a thin layer of oil on the hydrophobic surface for the microtubules to sediment to. The creation of the nematic was aided by centrifuging for 5 minutes at 1000 RPM (Sorvall Legend RT #6434). Images were acquired on an inverted Nikon Ti Eclipse with an Andor Clara camera using LC-PolScope microscopy.

**Computation of the defect core size:** We define defect cores as contiguous regions having the value of scalar order parameter $S \leq 0.5 S_{max}$ (see Fig. A1). These regions were extracted from the data using a floodfill algorithm. The scalar order parameter $S$ was obtained from the Q-tensor: $S = \sqrt{2\,Tr(Q^2)} = 2\sqrt{Q_{xx}^2 + Q_{xy}^2}$. In the simulations, the Q-tensor was directly available. In experiments, the Q-tensor was computed by coarse graining the molecular tensor obtained from the measured orientation field. We define the defect core size $d$ as the diameter of the defect core area averaged over the sample $\langle A \rangle$, assuming a circular core: $\langle d \rangle = \sqrt{4\langle A \rangle/\pi}$.

**Characteristic time:** We define the characteristic time $t_c$ as the time it takes for a +½ defect to travel a distance equal to its core size. The defect velocity is known to be proportional to $\alpha R/\eta$ [24], where $R$ is the core size, and thus $t_c \sim \eta/\alpha$. In both simulations and experiments, we compute the defect core size by finding the average area of the contiguous regions that have the scalar order parameter $S \leq 0.5\, S_{max}$ (See Appendix section "Computation of the defect core size"). The defect velocities are computed by tracking +½ defects over their lifetimes. The average defect diameter in the experiments is found to be 12 microns, with defect speed ~4 um/sec. This gives $t_c = $ 3 sec. The framerate of the data is 0.5 frames / second. This means that one frame $\approx 0.7\, t_c$ in experiments. A similar calculation for the simulation data yields that one simulation frame is rougly $1.4\, t_c$. The defects last for ~14 $t_c$ in the experiments before annihilating / moving out of the field of view, and ~ 35 $t_c$ in the simulations.

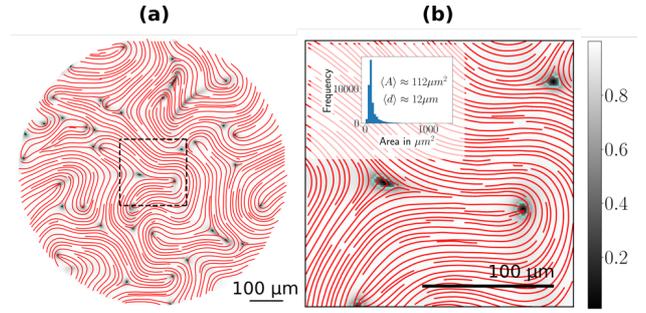

**Figure A1.** Illustration of the measurement of defect core size from experimental data. (a) Orientation field (red lines) of a sample frame superimposed on a heatmap of the computed scalar order parameter $S$. (b) Zoomed-in plot of the dashed region from the left panel. The color bar is the same for both the panels. Defect cores are defined as contiguous regions with $S \leq 0.5 S_{max}$, shown here with the green boundaries. (*Inset*) Histogram of the defect core areas. The average defect core size $d$ is defined as the diameter associated with the mean core area $\langle A \rangle$, assuming a circular core, $\langle d \rangle = \sqrt{4\langle A \rangle/\pi}$.

**Movie S1 (S1.gif):** This movie demonstrates an example of the forecasted defect nucleation process. *Left column*: the first eight frames are the input orientation fields. The forecasting results start at the 9[th] frame. *Middle column*: the ground truth retardance image sequence. *Right column*: the ground truth orientation field sequence. *Top row*: the whole view of the model. *Bottom row*: zoom in to the rectangle area in the top row to highlight the defect nucleation event. The green tricuspoids and red arrows indicate the forecasted –½ and +½ defects, respectively. The scale bar is 100 μm.

**Movie S2 (S2.gif):** This movie demostates an example of the forecasted defect anihilation process. *Left column*: the first eight frames are the input orientation fields. The forecasting results start at the 9[th] frame. *Middle column*: the ground truth retardance image sequence. *Right column*: the ground truth orientation field sequence. *Top row*: the whole view of the model. *Bottom row*: zoom in to the rectangle area in the top row to highlight the defect anihilation event. The green tricuspoids and red arrows indicate the forecasted –½ and +½ defects, respectively. The scale bar is 100 μm.

**Movie S3 (S3.gif):** This movie shows an example of the defect splitting process. *Left column*: the first eight frames are the input orientation fields. The forecasting results start at the 9[th] frame. *Middle column*: the ground truth retardance image sequence. *Right column*: the ground truth orientation field sequence. *Top row*: the whole view of the model. *Bottom row*: zoom in to the rectangle area in the top row

to highlight the defect splitting event. The red arrows indicate the forecasted +½ defects, respectively. The scale bar is 100 μm.

**Movie S4 (S4.gif)**: This movie shows an example of the temporal-spatial accumulation and propogation of forecasting errors. ***Left column***: the ground truth orientation fields. ***Middle column***: the forecasted orientation fields. ***Right column***: the corresponding temporal-spatial distributions of the forecasting fiedlity, which is calculated as $\cos(2\gamma)$, where $\gamma$ is the absolute angular difference between the ground truth and the forecasting result in a location. The green tricuspoids and red arrows indicate the forecasted –½ and +½ defects, respectively. The scale bar is 100 μm.

**Detection of defects in orientation fields**: The signed winding number $\frac{1}{2\pi}\oint \frac{\partial \theta}{\partial x} \cdot dx$, where $\theta$ is the location orientation, is calculated for every location in an orientation field using a pre-defined window-size [69]. The winding number is zero everywhere except at singular points [15, 17].

**Hydrodynamic theory of active nematics**
We use a simplified form of the Beris-Edwards equations [70, 71] for nematic hydrodynamics, extended to include active stresses [1, 2, 23-26, 28, 35-37, 72-74]. The hydrodynamic fields are the velocity $\vec{u}$ and the second rank tensor order parameter $Q$ with components $Q_{ij}$, with their dynamical equations of motion being
$$\partial_t Q + \vec{u} \cdot \nabla Q = (\Omega \cdot Q - Q \cdot \Omega) + \lambda E + D_r H$$
where $\Omega = \frac{1}{2}(\nabla u - (\nabla u)^T)$, $E = \frac{1}{2}(\nabla u + (\nabla u)^T)$, and $\lambda$ is the flow alignment parameter. Lastly, $H = -\delta F/\delta Q$, with
$$F = \frac{a_2}{2}Q^2 + \frac{a_4}{4}Q^4 + \frac{K}{2}(\nabla Q)^2$$
being the free energy of a passive nematic under the one-constant approximation [70], and $D_r$ is the coefficient of rotational diffusion. $a_2 = (1 - \rho)$ and $a_4 = (1 + \rho)/\rho^2$ are parameters responsible for bulk orientational order for densities $\rho > 1$, and K is the elastic modulus of the nematic.

The dynamics of the fluid are assumed to be incompressible, and are solved in the Stokes limit, with viscous forces balanced by substrate friction and active stresses:
$$\eta \nabla^2 \vec{u} - \Gamma \vec{u} - \nabla \cdot (\alpha Q) = 0$$
$$\nabla \cdot \vec{u} = 0$$
Here, $\eta$ is the viscosity, $\Gamma$ is the strength of the substrate friction, and $\alpha > 0$ is the strength of the extensile activity. These equations are solved in a large square domain with periodic boundary conditions. To ensure numerical stability, we use a semi-implicit finite-difference time-stepping scheme based on a convex splitting of the nematic free energy [75]. To solve the Stokes equation with incompressibility, we implement a Vanka-type box-smoothing algorithm on a staggered grid [76]. The solution at each time step is found using Gauss-Seidel relaxation iterations, and the rate of convergence to the solution is accelerated by using a multigrid method. The simulation codes are all in-house and are written in C. In non-dimensional units, we set K = 1, $\eta = 1$ and $D_r = 1$. For the simulations, we use $\rho = 1.3$, $\lambda = 1$, $\alpha = 0.2$ and $\Gamma = 0.03$.

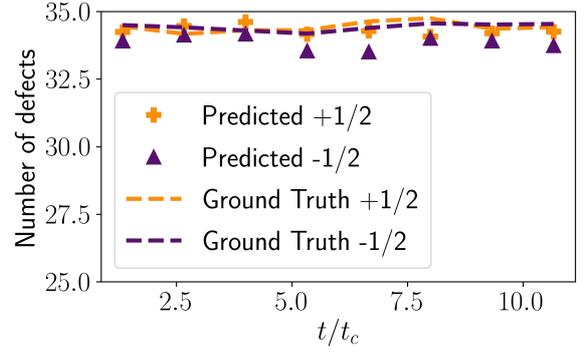

**Figure A2**. The overall performance of defect forecasting on the simulation data. Plotted are the number of defects in the viewed region that are detected in the data and forecasted by the model as a function of time, in units of the characteristic time, $t_c$. The number of defects in the forecasted frames only slightly decreases from the ground truth for the simulation data, as opposed to the larger decay observed in the prediction for the experimental data (See Fig. 5).